\documentclass{elsart}

\usepackage{amsmath,amssymb,epsf,latexsym,graphicx,makeidx,citesort}

\newcommand{\post}[2]{
\centering \leavevmode
 \includegraphics[width=#2cm]{#1}
 }

\newenvironment{proof}{\indent {\bf Proof.}}{\hfill \rule{.75em}{.75em}\bigskip}
\newenvironment{proofof}{\indent {\bf Proof of}}{\hfill \rule{.75em}{.75em}\bigskip}

\newcommand{\IR}{\mathbb{R}}

\begin{document}

\begin{frontmatter}

\title{Substitute Valuations: Generation and Structure}
\author{Bruce Hajek}


\begin{abstract}
Substitute valuations (in some contexts called gross substitute valuations)
are prominent in combinatorial auction theory.   An algorithm is given in this paper
for  generating a substitute valuation through Monte Carlo simulation.
In addition, the geometry of the set of all substitute
valuations for a fixed number of goods $K$ is investigated.   The set consists of a union
of polyhedrons, and the maximal polyhedrons are identified for $K=4$.  It is shown
that the maximum dimension of the polyhedrons increases with $K$
nearly as fast as two to the power $K$.    Consequently, under broad conditions,
if a combinatorial algorithm can present an arbitrary substitute valuation
given a list of input numbers, the list must grow nearly as fast as two to the power $K$.

\end{abstract}


\end{frontmatter}

\section{Introduction}   \label{sec.introduction}

Roughly speaking, one commodity is a substitute for another if the commodities are approximately
interchangeable.    In the economics literature, the notion of substitute valuations dates back to the
work of Walras on equilibrium theory and is prominent in the development of general equilibrium theory
(see, for example,  \cite{Wald51,ArrowDebreu54,ArrowBlockHurwiczI59}).
Kelso and Crawford \cite{KelsoCrawford82} formulated a version of substitute property for discrete
goods, opening the doors to a generalization of the theory of pricing and ascending auctions
that had been developed earlier for matching markets by Damange, Gale, Leanord, Shapley, Shubik,
Sotomayor, and others (see \cite{DemangeGaleSotomayor86}).   
More recently, new characterizations of substitute valuations have been found,  and algorithms and auctions for finding efficient allocations and market clearing prices for economies with buyers having substitute valuations have been found
\cite{GulStacchetti99,Ausubel05,Ausubel06,LehmannLehmannNisan05,HatfieldMilgrom05,MilgromStrulovici06d,ReijnierseGellekomPotters02,LienYan07}.

For readers unfamiliar with the concept of substitute valuations in economics,
we introduce it by describing a special case of an auction algorithm given in
\cite{KelsoCrawford82}.   Suppose there are $K$ goods to
be auctioned to $n$ buyers, using an ascending price auction with nonnegative
integer prices, as follows.  Initially the price for each good is zero and each good is
provisionally assigned to some buyer.  During each round of the auction, suppose
that if a good $k$ is not provisionally assigned to a particular buyer, then the buyer
can place a bid for the good at price $p_k+1$, where $p_k$ is the current price.
 If there are any bids made for a good in the round,
then the good is provisionally assigned to one of the buyers placing a bid for the
good, and the price of the good is increased to the price bid.    If no new bids are
made in a particular round, then the auction ends. 
Once a good is provisionally assigned to a buyer, if no higher bids are ever placed on
that good, it is sold to the highest buyer at the price bid.   Although all the goods are sold in
such an auction, the outcome can be very inefficient.   For example,
suppose one of the buyers would greatly value receiving two particular goods, but
that neither good by itself would be of any value to the buyer.  For example, the goods could be
two communication links in series.   The buyer could
place bids for both goods.   But suppose a second buyer aggressively competes for one
of the two goods, eventually outbidding the first buyer for that good.   Then the first buyer
could be stuck buying the other good, even though that good alone has no value to the buyer.
If all buyers have substitute valuations, however, and if all bid in a straight-forward manner
based on those valuations, then the auction is indeed efficient.   The substitutes
property is that, if a buyer prefers a particular bundle of goods for one set of prices,
and then if the prices of some goods are increased, then there is a new bundle preferred by the
buyer which includes all items in the original bundle that did not have a price increase.
Hence, if the valuations of all buyers satisfy the substitute condition, the vector of final prices
for the goods is such that the supply of goods (i.e. one of each type)
is matched by the demand.  In summary, the substitute valuation property is precisely
what is needed to make ascending price auctions efficient.   

Auctions have been implemented for sale of such diverse resources as: wireless spectrum licenses,
gate access at airport terminals, truckload transportation, bus routes, and polution permits \cite{deVriesVohra03,CramtonShohamSteinberg06}.
Large sums of money can be involved and it may be very expensive to rely on learning from experience.
There is thus a need to be able to produce simulations with buyers having realistic valuations.
Design of realistic models for valuations is an art that involves spatial and/or temporal dependencies
among the goods and buyers, which depend heavily on the particular market addressed.
This topic is beyond the scope of this paper, but we refer to \cite{Leyton-BrownShoham06} for
background, and presentation of a graph based approach to modeling dependencies.

Substitute valuations (a.k.a. gross substitute valuations) play a central role in the
theory of auctions because they (1)  are in a sense necessary for
existence of market clearing prices \cite{GulStacchetti99,Milgrom00},  (2) are related to existence of
monotone price auctions, (3) are related to desirable monotonicity of prices and immunity
from strategic behavior by subgroups of buyers for the important family of Vickrey auctions
\cite{AusubelMilgrom06}.    In practice, the valuations of buyers in a given
auction might not have the substitutes property.   However, designing algorithms that work well in
particular for substitute valuations, but also have reasonable behavior for the other valuations one
is likely to encounter in a given market, is a reasonable approach to practical auction design
\cite{Milgrom04}.

Given the importance of the class of substitute valuations, it is useful to be able to:
\begin{itemize}
\item Check whether a given valuation $v$ has the substitute property.
\item  Generate random substitute valuations by Monte Carlo simulation,
for the purposes of testing auction algorithms and exploring the structure of substitute valuations.
\item  Determine the size of the class of substitute valuations, in various senses, for example
to determine how much information must be generated, transmitted, or stored in
connection with the use of the valuations.
\end{itemize}
There exist nice results addressing the first or these items, including the result of
 \cite{ReijnierseGellekomPotters02,LienYan07} given as Proposition \ref{prop.S3}
 below, and the connection to matroid theory mentioned at the end of the next section.
 Also, a valuation has the substitute property if  and only if its dual is
 submodular \cite[Theorem 10]{AusubelMilgrom02}.
 
The contributions of this paper are focused on the second and third items.
First, a complete parameterization of all (nondecreasing) substitute
valuations on four items is given in Section \ref{sec.4goods}.   Our motivation for this
is threefold:  (1) Gain intuition on the family of all substitute valuations,
(2) Illustrate the geometry of the space of substitute valuations as a union of maximal polyhedrons, and identify the
dimension of the polyhedrons.  These concepts are used in Section 6.  
(3)  Point out a lemma that will be used in the proof of convergence of the Monte Carlo algorithm.

Second,  a Monte Carlo simulation algorithm for
generating substitute valuations is given in Section \ref{sec.MonteCarlo}. 
The first step of the algorithm is to generate an arbitrary valuation, and then the algorithm
performs a finite number of modifications leading to a substitute valuation.   A proof of
convergence is given which involves the characterization of a substitute valuation on the
class of sets of a given size by a local exchange property.   

The third contribution of the paper is to give an indication of how much richer the class of substitute
valuations is than the subclass arising from the assignment problem.
Section  \ref{sec.assign_val} shows that the assignment valuations comprise a useful and interesting
set  of substitute valuations, which covers much of the set of all substitute valuations in the
case of four goods.   In particular, a subclass of assignment allocations is reviewed
in  Section \ref{sec.mono_assign} for which essentially no computation is needed to compute
the value of a given bundle.  This section gives further intuition about the use of the
dimension of the space of allocation valuations leading to a better understanding of the result of Section 6.
Section \ref{sec.four_assign} examines how much of the space of substitute valuations is covered by the
assignment valuations  for the case of four goods.   This naturally ties together Sections  \ref{sec.4goods} and \ref{sec.mono_assign}.

Fourth, Section  \ref{sec.spec_val} displays a set of substitute valuations that, for a large number of
goods, is markedly different from the set of
assignment valuations.   The existence of those valuations implies that for a large class
of algorithms for presenting substitute valuations, the number of real-valued
inputs must grow exponentially in the number of goods.
Section \ref{sec.spec_val} shows that for any $\epsilon > 0$, if $K$ is sufficiently large, then the space of substitute valuations contains polyhedrons with dimension at least $2^{(1-\epsilon) K}$.   As explained, this has a
negative implication regarding the existence of presentation algorithms capable of producing arbitrary
substitute valuations.

The results of this paper are interrelated as follows.
As mentioned above, we present an algorithm for generating a substitute valuation
in Section  \ref{sec.MonteCarlo}.  A related problem is to find algorithms
for {\em presenting} substitute valuations.   The number of values of a valuation, $2^K$, grows quickly
with $K$.   The idea of a presentation algorithm is to ask a buyer to specify a smaller list of numbers,
and to use a polynomial complexity algorithm which, given the input list from the buyer and a bundle
of goods,  can determine the value of the bundle.    The use of allocation valuations, discussed in
Section \ref{sec.assign_val}, is an excellent prototype of a presentation algorithm. 
For allocation valuations, the list of numbers specified by a buyer are the weights on a bipartite
graph, and the value of a bundle of goods is given by the maximum weight of the matchings
in a subgraph determined by the bundle.
More complex but related presentation algorithms are given in \cite{BingLehmannMilgrom04}.  
However, there are no known presentation algorithms which cover the entire set of substitute valuations.
If such an algorithm did exist, one could use it to generate random valuations by inputting to the
algorithm a list of random numbers, offering a perhaps more attractive alternative to our generation
algorithm in Section  \ref{sec.MonteCarlo}.    Thus, the question of whether there exists a presentation
algorithm that can cover the set of all substitute valuations is naturally related to the problem of
simulating substitute valuations.    This question is addressed in this paper in two ways:  First, by giving
the structure of the space of all substitute valuations for four goods.  Secondly, by showing (Section  \ref{sec.spec_val})
that, for large numbers of goods, under fairly general conditions, for any $\epsilon > 0$, the list of numbers a buyer
would need to input to specify
arbitrary substitute valuations must have length at least $2^{(1-\epsilon)K}$ for large $K$.  This negative
result about possibilities for presentation algorithms strengthens the case to search for good generation
algorithms.  A final connection between sections  is that we used the algorithm of Section \ref{sec.MonteCarlo}
to help discover the class of valuations presented in Section \ref{sec.spec_val}.

Section \ref{sec:background} gives notation and background,  states some useful known properties
of substitute valuations which can be used to build up families of such valuations,
and briefly points to where substitute valuations can be found under a different name
in some recent literature on matroid theory.

\section{Notation and background}  \label{sec:background}

Suppose $K$ goods, represented by ${\cal K}=\{1, \ldots, K\}$, are to be
allocated.   The set of possible bundles of goods (i.e. subsets of $\cal K$)
is denoted by $2^{\cal K}$.  If $A$ is a bundle and $i$ is a good not in $A$, we write $Ai$ for the
set $A \cup i$.   If $A$ is a bundle and $i$ and $j$ are distinct goods not in
$A$, we write $Aij$ for $A \cup \{i,j\}$.   Similarly, if $i$ and $j$ are distinct goods,
we sometimes write $ij$ to denote the set $\{i, j\}$.  The notation $|A|$ denotes the
cardinality of $A$.

A function $f : 2^{\cal K} \rightarrow \IR$ is {\em submodular} if for any bundles $A$ and $B$,
$f(A\cup B)-f(A)-f(B)+f(A\cap B) \leq 0$.   An equivalent condition is that
$f(Aij)-f(Ai)-f(Aj)+f(A) \leq 0$, whenever $A$ is a bundle and $i$ and $j$ are goods not in $A$
(see \cite{Topkis98}).   A function $f$ is {\em supermodular} if $-f$ is submodular.

\begin{defn}
A triplet of numbers $(x,y,z)$ has the {\em double maximum property} if
 at least two of the numbers are equal to the maximum of the three numbers, or
 equivalently, $x\leq \max\{ y , z \}$ and  $y\leq \max\{ x, z \}$ and  $z\leq \max\{ x,y \}$.
 A triplet of numbers $(x,y,z)$ has the {\em double minimum property} if
 at least two of the numbers are equal to the minimum of the three numbers,
 or equivalently,  if  $x\geq \min\{ y , z \}$ and  $y\geq \min\{ x, z \}$ and  $z\geq \min\{ x,y \}$.
 \end{defn}

Following terminology common in the economics literature, a {\em valuation}
$v$ is a mapping from $2^{\cal K}$ to $\IR$.    Throughout this paper, we
require that valuations be normalized to assign value zero to the
empty set.  We consider quasi-linear payoffs.      
Thus, given a price vector $p$, by which we mean an element of $\IR_+^K$,  the payoff  function of a
buyer with valuation $v$ is $v(A)-p\cdot A$ for $A \in 2^{\cal K}$, where the notation
$p\cdot A = \sum_{k\in A} p_k$ is used.
The demand correspondence $D$ for a buyer with valuation $v$ is defined by
$ D(p) =  \arg\max_A  v(A)-p\cdot A.$ 

\begin{defn}  \label{def:sub} (Substitute valuation \cite{KelsoCrawford82})  A valuation $v$ for $K$
distinct goods is a {\em substitute valuation} if, for any price vectors $p$ and $q$ such
that $p\leq q$ and any $A \in D(p)$, there exists a bundle $A' \in D(q)$ such
that $\{k \in A : p_k=q_k\} \subset A'$.
\end{defn}

The following condition is a key to checking whether a given valuation has
the substitute property.  (A brief intuitive explanation is given after
Fact \ref{prop.v_to_theta} below.)

\begin{defn} (Properties $S3(L)$ and $S3$)   Let $2  \leq L \leq K-1$.   A valuation $v$ has
property $S3(L)$ if
$(v(Aij)+v(Ak), v(Aik) + v(Aj), v(Ajk) +v (Ai) )$ has the double maximum property
whenever $A$ is a bundle with $|A|=L-2$, and $i,j,k$ are distinct goods not in $A$.
The valuation $v$ is said to satisfy $S3$  if it satisfies $S3(L)$ for $2 \leq L \leq K-1.$
\end{defn}

\begin{prop}  \cite{ReijnierseGellekomPotters02,LienYan07}  \label{prop.S3}
A nondecreasing valuation $v$ is a substitute valuation if and only if
$v$ is submodular and $v$  has property $S3$.
\end{prop}

It is useful to represent a valuation as a linear function minus an
{\em interaction function,} defined as follows.
\begin{defn}  An interaction function is a function
$\theta : 2^{\cal K} \rightarrow \IR$ such that  $\theta(A) = 0$ for bundles $A$ with $|A| \leq 1$.
\end{defn} 
Given a valuation  $v$, let $\mu$ be the vector of valuations of singleton sets,
so $\mu(k)=v(\{k\})$ for  $1\leq k \leq K$.   Then the {\em interaction function of $v$} is defined by
$\theta(A)=\mu \cdot A - v(A)$ for all bundles $A$.  
Obviously, $v(A)=\mu\cdot A - \theta(A)$.   In order to specify a substitute
valuation $v$, it suffices to specify $\mu$ and $\theta$.   
We work with the interaction function rather than always working with the original valuation
in this paper mainly for two reasons.
 First, in the case of four goods (K=4, Sections  \ref{sec.4goods} and \ref{sec.four_assign}), it is
easier to deal  with the 11 nonzero values of the interaction function than with the
15 nonzero values of the original valuation. 
Second, in the generation algorithm we present in Section \ref{sec.MonteCarlo}, the sequence of interaction functions
converging to the final output is monotone nondecreasing.  This could be written
in terms of the original valuations, but then there would be difficulty with
the valuations going negative or not being monotone nondecreasing.
The properties $S3$ and $S3(L)$ can be expressed in terms of $\theta$ or in terms of 
another related function, as described next.

\begin{defn} (Properties $S3_{\theta}(L)$ and $S3_{\theta}$)   Let $2 \leq L \leq K-1$.   An
interaction function $\theta$ has property $S3_{\theta}(L)$ if
$(\theta(Aij)+\theta(Ak) , \theta(Aik) + \theta(Aj), \theta(Ajk)+\theta(Ai) )$ has the
double minimum property
whenever $A$ is a bundle with $|A|=L-2$, and $i,j,k$ are distinct goods not in $A$.
An interaction function $\theta$ is said to satisfy $S3_{\theta}$  if it satisfies $S3_{\theta}(L)$
for $2\leq L \leq K-1$.
\end{defn}

The {\em two-point conditional interaction function} $\delta$ for a valuation $v$ is defined as follows.
For any bundle $A$ and goods $i$ and $j$ not contained in $A$, 
$\delta_{ij|A}=( v(Ai) - v(A) )  + (v(Aj) -v(A))  - ( v(Aij) - v(A) ),$
so that, intuitively, $\delta_{ij|A}$ is the penalty in value for the buyer acquiring both $i$ and $j$, given
the buyer has already acquired the bundle  $A$.  The definition
simplifies to $\delta_{ij|A}= v(Ai)  + v(Aj) - v(Aij) -  v(A),$ or it can be written in terms
of $\theta$ as   $\delta_{ij|A}= \theta(Aij)  -  \theta(Ai)  - \theta(Aj)  + \theta(A)$. 
For brevity we write $\delta_{ij}$ instead of $\delta_{ij|\emptyset}.$  Note that
$\delta_{ij}=\theta(ij)$ for distinct goods $i$ and $j$. 

\begin{defn} (Properties $S3_{\delta}(L)$ and $S3_{\delta}$)   Let $2 \leq L \leq K-1$.   A
two-point conditional interaction function $\delta$  has property $S3_{\delta}(L)$ if
$( \delta_{ij|A} , \delta_{ik|A} ,  \delta_{jk|A} )$
has the double minimum property
whenever $A$ is a bundle with $|A|=L-2$, and $i,j,k$ are distinct goods not in $A$.
The function $\delta$ is said to satisfy $S3_{\delta}$  if it satisfies $S3_{\delta}(L)$
for $2\leq L \leq K-1$.
\end{defn}

The following facts are obvious.

 \begin{fact}   \label{prop.v_to_theta}
 If $v$ is a valuation with interaction function $\theta$ and two-point conditional interaction function
 $\delta$, then: \\
(a) The following are equivalent: $v$ is submodular, $\theta$ is supermodular,  $\delta$ is nonnegative. \\
(b)  $v$ is nondecreasing if and only if $\mu(k) \geq \max_{A:k\not\in A} \theta(Ak)-\theta(A)$ for all $k$.\\
(c) For $2\leq L \leq K-1$,  the following are equivalent:  $v$ satisfies $S3(L)$, $\theta$ satisfies
$S3_{\theta}(L)$, $\delta$ satisfies $S3_{\delta}(L).$   \\
(d)   The following are equivalent:  $v$ satisfies $S3$, $\theta$ satisfies
$S3_{\theta}$, $\delta$ satisfies $S3_{\delta}.$   \\
 \end{fact}
 
 For the reader unfamiliar with Proposition \ref{prop.S3}, we give a brief explanation for why substitute valuations
 must be submodular (we'll show for $K=2$) and why they must satisfy condition $S3$  (we'll show
 $S3_{\delta}(2)$ holds for $K=3$).    See \cite{LienYan07} for a complete, direct proof of the general result
 of Proposition \ref{prop.S3}.
 If $K=2$ and $v(ij) > v(i) + v(j)$ (violating submodularity) then prices
 $p_i$ and $p_j$ could be selected so that $v(ij) > p_i + p_j$, $v(i)< p_i$, and $v(j)< p_j$.
 Then $\{i,j\} \in D(p)$.  But if $p_i$ is increased enough, the demand set shrinks to $\emptyset$, instead
 of including $j$, so $v$ is not a substitute valuation.
 Moving to $K=3$, suppose $v$ is submodular, but that the condition
$\delta_{ij} \geq \min\{\delta_{ik}, \delta_{jk}\}$ fails to hold.
Then the price vector $p$ can be selected so that
$$
0\leq \delta_{ij} ~<~ (v(j)-p_j)~<~(v(i)-p_i)~=~(v(k)-p_k)~<~\min\{ \delta_{ik}, \delta_{jk} \}.$$
Then $\{i,j\} \in D(p)$.  Indeed, $\{i\}$ yields the same payoff as $\{k\}$ and a greater
payoff than either  $\{j\}$ or $\emptyset$.
Since $\delta_{ij} < v(j)-p_j$, $\{i,j\}$ has a larger payoff than $\{i\}$.   By the same reasoning,
$\{i,k\}$ and $\{j,k\}$ have smaller payoffs than $\{k\}$, or equivalently, $\{i\}$. 
So $\{i,j\}$ has a larger payoff than any other bundle of zero, one, or two goods. 
Finally, by submodularity, the change in payoff for adding
$k$ to $\{i,j\}$ is less than or equal to the change in payoff for adding $k$ to $\{j\}$ alone, which is
negative.  Thus, $\{i,j\} \in D(p)$.  But if $p_i$ is greatly increased, the unique new demand set is $\{k\}$,
which does not include $j$ as required by the substitute condition.

Next, we collect together some facts about substitute valuations
 which are useful for building up interesting classes of valuations.   For example,
 these properties are used in the construction of presentation algorithms in \cite{BingLehmannMilgrom04}.
 A valuation $v$ is {\em linear} if $v(A)=\mu\cdot A$ for a vector $\mu$ of nonnegative
weights.   A valuation $v$ is {\em additively
concave} if there is a partition of $\cal K$ into (disjoint) subsets $S_1, \ldots , S_J,$
and if there are nonnegative concave functions $\phi_1, \cdots , \phi_J,$ such that
$v(A)= \sum_j  \phi_j (|A\cap S_j| )$.
Both linear valuations and additively concave valuations are substitute valuations, as
can be seen directly from the definition of substitute valuations
\cite{GulStacchetti99}.

The {\em aggregate}, or max convolution, of two valuations, $v_1$ and $v_2$, is
the valuation $v_1 * v_2$,  defined by:
$$
v_1  *  v_2 (A) = \max_{B\subset A}  v_1(A-B) + v_2(B).
$$
The value $v_1*v_2(A)$ is the sum of values for two buyers if the bundle $A$ is
optimally split between them.
The family of substitute valuations is closed under aggregation
\cite{LehmannLehmannNisan05,MurotaShioura99}.
A valuation $v$ is called a {\em single unit valuation} if there exist nonnegative
weights $(w_{k} : k\in {\cal K} )$ such that $v(A)=\max \{ w_{k} : k \in A \}$.    Single unit valuations
are substitute valuations.   Section \ref{sec.assign_val} discusses assignment
valuations, which arise as the aggregation of multiple single unit valuations.

Following Gul and Stacchetti \cite{GulStacchetti99},  given $L$ with $0 \leq L \leq K$,
the {\em L-satiation} of a valuation $v$ is the valuation $\widehat{v}$ defined by
$$
\widehat{v}(A)=\max_{B\subset A, |B| \leq L }  v(B).
$$
The $L$-satiation of a substitute valuation is also a substitute valuation.
This result was obtained by Gul and Stacchetti \cite{GulStacchetti99} for
linear or additively concave valuations, and in general by Bing et al.
 \cite{BingLehmannMilgrom04}. 
 Another proof is given below (see Corollary \ref{cor.LSAT}).

For completeness, we point out that the class of substitute valuations forms a natural link
 between the theory of matroids and auction theory.   With the exception of
 Remark \ref{remark.Mconcave}, the terminology
 in this paragraph is not used elsewhere in this paper.
 Fujishige and Yang \cite{FujishigeYang03} showed that, within the class of monotone valuations,
the class of substitute valuations is equivalent to the class of $M^\natural$-concave (read ``$M$ natural concave")
functions introduced by Murota and Shioura  \cite{MurotaShioura99}.   The notion
of $M^\natural$-concavity is an extension of the notion of $M$-concavity, introduced
by Murota \cite{Murota96se}.  In turn, $M$-concavity is a generalization of the notion of
valuated matroids introduced by Dress and Wenzel  \cite{DressWenzel90,DressWenzel93}, and
valuated matroids are generalizations of matroid rank functions.
As pointed out by Gale \cite{Gale68}, matroids are intertwined with the theory of problems for which
the greedy algorithm is optimal.    Substitute valuations are associated with markets such that
a simple ascending price auction is optimal.    Furthermore, the assignment problem (a.k.a. weighted
bipartite matching problem) is a prototype of both the theory of algorithms in matroid theory,  and
equilibrium theory in economics.   So the connection between matroid theory and auction theory is a strong one.

\section{Parameterization of substitute valuations on four goods}  \label{sec.4goods}

 Let ${\cal S}_K$ denote the set of all nondecreasing substitute valuations
on $\cal K$ (with value zero at $\emptyset$), viewed as a subset of $\IR^{2^{\cal K}}.$
Recall that a {\em polyhedron} (or polyhedral set)  is a nonempty set that can be represented as the intersection
of finitely many half-spaces.   The following is a corollary of Proposition \ref{prop.S3}:

\begin{cor}  \label{cor.polyhedral_structure}
For $K\geq 1$, the set ${\cal S}_K$ can be represented as the union of
finitely many polyhedrons.
\end{cor}
 
\begin{proof}
By Proposition  \ref{prop.S3}, the set ${\cal S}_K$ is the  subset of $\IR^{2^{\cal K}},$ satisfying the
normalization at $\emptyset$, monotonicity,  submodularity, and $S3$  conditions.   The normalization
constraint, namely $v(\emptyset)=0$, requires that $v$ be in the intersection of the two half-spaces,
$\{v(\emptyset)\leq 0\}$ and $\{v(\emptyset) \geq 0\}$.
Also, each constraint in the definition of monotonicity or submodularity
is equivalent to constraining $v$ to be in a half-space.     Condition $S3$  is
equivalent to the requirement that for any bundle $A$ and ordered set of goods $i,j,k$ not in $A$, at least
one of the following two constraints holds:
$$
v(Aij)+v(Ak) \geq v(Aik)+v(Aj)  ~~\mbox{or}~~ v(Aij)+v(Ak) \geq v(Ajk)+v(Ai).
$$
That is, $v$ must be in one of two half-spaces.  Making a particular
choice of half-space for each such $A,i,j,k$, thus specifies a subset of substitute valuations
forming a polyhedron.   The union of such polyhedrons, over all choices for the half-space
for each $A,i,j,k$, is ${\cal S}_K$.
\end{proof}
 
The {\em maximal polyhedral subsets} of ${\cal S}_K$ are the polyhedral subsets of
${\cal S}_K$ which are not proper subsets of any other polyhedral subsets of ${\cal S}_K$.
Corollary \ref{cor.polyhedral_structure} implies that ${\cal S}_K$ is equal to the union
of its maximal polyhedral subsets.
The {\em dimension} of a polyhedron is the dimension of the smallest affine subspace containing the polyhedron.
In this section we identify the maximal polyhedral subsets of ${\cal S}_K$ for
$1\leq K \leq 4$, and note their dimensions. The emphasis is on the case $K=4$, but the other cases help
build intuition.

\subsection{$K$=1}
Clearly, ${\cal S}_1=\{  (v(\emptyset), v(1)) : v(\emptyset)=0, v(1) \geq 0 \}$, so that
${\cal S}_1$ itself is a polyhedron, which is one dimensional.

\subsection{$K$=2}
The set ${\cal S}_2$ consists of all four-vectors 
$(v(\emptyset), v(1),v(2),v(12) )$ satisfying the normalization constraint: $v(\emptyset)=0$,
the monotonicity constraints:  $v(1) \geq 0, v(2) \geq 0, v(12)\geq v(1), v(12)\geq v(2),$
and the submodularity constraint  $v(12)-v(1)-v(2)+v(\emptyset)\leq 0.$     Note that
condition S3 is vacuous for $K=2$.
Therefore, ${\cal S}_2$ itself is a polyhedron, which is three dimensional.

\subsection{$K$=3}
The set ${\cal S}_3$ consists of vectors of length eight, so it is tedious to write
out the constraints directly in terms of the values of $v$.    We instead use
the representation $v(A)=\mu\cdot A - \theta(A)$.    Let $v$ be a substitute valuation on ${\cal K}=\{1,2,3\}$.
Let $a=\min\{\theta(12),\theta(13),\theta(23) \}$.
By the double minimum property of $\theta$, $S3_{\theta}$,  there is a permutation $(i,j,k)$ of $(1,2,3)$
so that  $\theta(ij)=\theta(ik)=a. $ Let $b=\theta(jk)$ and $c=\theta(ijk)$.   We claim that the following conditions
 are satisfied by the six parameters $a,b,c, \mu_i, \mu_j, \mu_k$:
\begin{equation}   \label{eq.K3}
 0\leq a \leq b,    ~~~  a+b \leq c,     ~~~  \mu_i \geq c-b, ~~~ \mu_j \geq c-a, ~~~       \mu_k \geq c-a.
\end{equation}
We have $ 0\leq a$ by the supermodularity of $\theta$ (or equivalently the  submodularity of $v$),
and $a\leq b$ by the choice of $a$ and $b$.
The next inequality, $a+b \leq c$, also follows from the supermodularity of $\theta$.
The last three inequalities in \eqref{eq.K3} result from the monotonicity of $v$: they insure that
adding a third good to the other two does not decrease $v$.   So any substitute valuation for $K=3$
can be represented as above as claimed.  Conversely, if $(i,j,k)$ is a permutation
of $(1,2,3)$ and the six parameters $a,b,c, \mu_i, \mu_j, \mu_k$ satisfy \eqref{eq.K3}, then the interaction
function $\theta$ with $\theta(ij)=\theta(ik)=a$, $\theta(jk)=b$, and $\theta(ijk)=c$, together with $\mu$,
determines a substitute valuation.   (Since submodularity is insured by the first two sets of inequalties, monotonicity
of $v$ when going from two goods to three implies monotonicity in general.)
The set of all substitute valuations obtained this way, for the permutation
$(i,j,k)$ fixed, specifies a six dimensional polyhedral subset of ${\cal S}_3$.   Goods $j$ and $k$ play
a symmetric role in the above, so that the same polyhedron results if $j$ and $k$ are swapped.
Thus, ${\cal S}$ is the union of three six-dimensional polyhedrons, determined as above for $i=1$, $i=2$, or $i=3$.
The three polyhedrons can also be compactly expressed as 
${\cal S}_3 \cap \{ \delta_{12}=\delta_{13}  \},  {\cal S}_3 \cap \{ \delta_{12}=\delta_{23}  \},$  and
${\cal S}_3 \cap \{ \delta_{13}=\delta_{23}  \} $.
The intersection of any two of these three maximal polyhedrons is the five-dimensional polyhedron
${\cal S}_3 \cap \{ \delta_{12}=\delta_{13}=\delta_{23} \}$.

\subsection{$K$=4}

Let $v$ be a valuation on ${\cal K}=\{1,2,3,4\}.$
We can focus on identifying the possible values of the interaction function $\theta(A)$
on bundles with $|A|=2$ and $|A|=3$.  Indeed, suppose $\theta$ is specified on such
sets consistently with supermodularity (i.e. so that $\delta_{ij}\geq 0$ and $\delta_{ij|k} \geq 0$
for distinct goods $i,j,k$).    Then,   $\theta$
 will be supermodular if $\theta(1234)$ is large enough, and
$v$ will be nondecreasing if the components of $\mu$ are large enough.   The resulting
$v$ will be a nondecreasing substitute valuation if and only if $\theta$
satisfies conditions $S3_{\theta}(2)$ and $S3_{\theta}(3)$, or equivalently, $\delta$ satisfies
conditions $S3_{\delta}(2)$ and $S3_{\delta}(3)$.

Suppose that $v$ is a substitute valuation.
Consider the complete undirected graph with vertex set $\cal K$
and associate with an edge $ij$ the value $\delta_{ij}$.  By condition $S3_{\delta}(2)$,
the $\delta$'s around any triangle in the graph have the double minimum property.
Let $a$ be the minimum value of the $\delta$'s for the six edges.
At least one vertex has two incident edges with $\delta$'s equal to $a$.  
Denote by Case 1  the case that some vertex  $i$ has three edges incident with
$\delta$'s equal to $a$.   In Case 1, let $b$ denote the smallest $\delta$ value
on the triangle $jkl$.   By permuting the vertices if necessary, we assume that
$\delta_{jk}=\delta_{jl}=b$.   Let $c=\delta_{kl}$.   Then $a \leq b \leq c$, and it is
possible that $a<b<c$.  Denote by Case 2 the case that there is a cycle of length four such that all four
$\delta$'s around the cycle are equal to $a$.  By renumbering the vertices, if necessary,
we can assume in Case 2 that $\delta_{ij}=\delta_{jk}=\delta_{kl}=\delta_{il}=a$.
Let $\delta_{ik}=b$ and $\delta_{jl}=c$.   Then $a \leq \min\{b,c\}$, and it is possible that
$a < \min\{b,c\}$.

\begin{figure}[htb]
\post{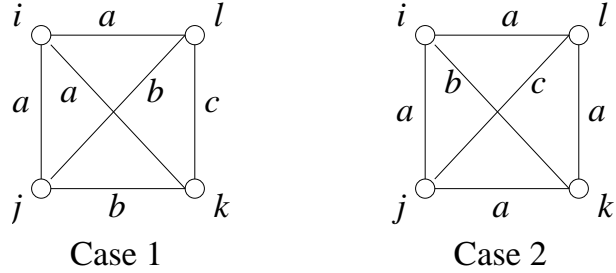}{8}   
 \caption{Possibilities for the $\delta$ values for a substitute valuation with $K=4$.  \label{fig.Sfour} }
 \end{figure}

The two cases are indicated in Figure \ref{fig.Sfour}.  On one hand, at least one of the
cases must hold (for a suitable labeling of the vertices).  On the other hand,
both cases hold if and only if at least five of the $\delta$'s are equal to $a$.  
By checking each of the two cases, the following lemma, used in the next section, is easily verified. 
\begin{lem}  \label{lemma.F4} If $v$ satisfies $S3_{\delta}(2)$ then 
$(\delta_{ij}+\delta_{kl}, \delta_{ik}+\delta_{jl}, \delta_{il}+\delta_{jk})$ has the double minimum property
for any distinct goods $i,j,k,$ and $l$.
\end{lem}
 
Condition $S3_{\delta}(3)$ is summarized in Table \ref{table.diffdelta}.  
Each row of the table lists a single good, followed by the three $\delta$'s given that good.
For example, the first row begins with good $i$, and the three values following it
are equal to $\delta_{kl|i}, \delta_{jl|i},$ and $\delta_{jk|i}$.
Here $\theta_{-i}=\theta_{jkl}=v(j)+v(k)+v(l)-v(jkl)$, and $\theta_{-j},\theta_{-k}$, and
 $\theta_{-l}$ are defined similarly, and we use the fact $\delta_{ij|k}=\theta_{-l}-\delta_{ik}-\delta_{jk}$.
Condition $S3_{\delta}(3)$ means that the three quantities in
each row of the table have the double minimum property. 
\begin{table}[htb]    \caption{Summary of Condition $S3_{\delta}(3)$ for $K=4$.  \label{table.diffdelta} }
 $$
 \begin{array}{c|cccc}
 i  ~&   ~    &  \theta_{-j} - \delta_{ik}  - \delta_{il}  ~ &  \theta_{-k}   - \delta_{ij}  - \delta_{il} ~ &  \theta_{-l}  - \delta_{ij}  - \delta_{ik}    \\
j  ~&    \theta_{-i}   - \delta_{jk}  - \delta_{jl} ~  &	    &  \theta_{-k}  - \delta_{ij}  - \delta_{jl}  ~ &  \theta_{-l}  - \delta_{ij}  - \delta_{jk}    \\
k  ~&    \theta_{-i}   - \delta_{jk}  - \delta_{kl}   ~&   \theta_{-j} - \delta_{ik}  - \delta_{kl}   ~ &      &  \theta_{-l}   - \delta_{ik}  - \delta_{jk}   \\
l  ~&    \theta_{-i}  - \delta_{jl}  - \delta_{kl}   ~ &	  \theta_{-j} - \delta_{il}  - \delta_{kl}  ~  &  \theta_{-k} - \delta_{il}  - \delta_{jl}   ~ &    
 \end{array}
 $$
\end{table}

 \begin{table}[h]
 \caption{Specialization of Table \ref{table.diffdelta}   to Case 1, with the main subcase indicated by boxes.  \label{table.diffdelta1} }
 $$
 \begin{array}{c|cccc}
 i  &       &	  \theta_{-j} - a  - a   & \framebox{$\theta_{-k}   - a  -a$} &  \framebox{$ \theta_{-l}  - a -a$}   \\
j  &    \theta_{-i}   - b -b  &	    & \framebox{$\theta_{-k}  - a  - b$}  & \framebox{$ \theta_{-l}  - a - b$}  \\
k &    \framebox{$\theta_{-i}   - b -c$}  &   \framebox{$\theta_{-j} -a - c $}  &      &  \theta_{-l}   - a  - b   \\
l  &    \framebox{$\theta_{-i}  - b  - c$}   &	\framebox{$  \theta_{-j} -a  - c $}  &  \theta_{-k} - a - b    &    
 \end{array}
 $$
 \end{table}
 
 Let us now examine Case 1 further.    If Case 1 holds, Table \ref{table.diffdelta} becomes Table \ref{table.diffdelta1}.  (Ignore the boxes in  Table \ref{table.diffdelta1} for a moment.)      It
turns out that if $a=b$ or $b=c$, or if the three entries in a row of the table are all equal, then
the substitute valuation $v$ is in the intersection of multiple maximal polyhedrons of ${\cal S}_4$.
So, for the purposes of identifying individual maximal polyhedrons, assume that $a < b < c,$  
and seek values of $\theta_{-i}$, $\theta_{-j}$, $\theta_{-k}$, and $\theta_{-l}$ so that precisely
two entries in each row achieve the row minimum.   {\em A priori}, there are 81 possible choices
of  two entries per row of Table  \ref{table.diffdelta1} to be row minimums, but as we will see,
the condition  $a < b < c$ greatly reduces the number of possibilities.   One possibility is indicated
by the boxes in Table \ref{table.diffdelta1},  showing which entries in each row are equal to the
minimum value in the row.    We call this {\em the main subcase of Case 1}.   The two boxed
terms in each row are equal, and the third term in each row is strictly larger than the two equal
terms, if and only if the following two conditions are satisfied: 
$\theta_{-k}=\theta_{-l}$, $\theta_{-i}=\theta_{-j}+b-a$,  and $a-b > \theta_{-k} - \theta_{-i}  > a-c$.
Looking for more possibilities, notice that if the last two entries in the first row are boxed as
in the main subcase of Case 1, then the same must be true in the second row.   Similarly,  the choice
in the third row forces the choice in the fourth row.  Suppose we change the choice in the third row
to boxing the second and third terms.  This forces $\theta_{-i}+a-b > \theta_{j} = \theta_{-l}+c-b$,
and therefore forces the second and third terms in the fourth row to be boxed.
This gives rise to the case shown in Table \ref{table.diffdelta1i}, in which no term involving
$\theta_{-i}$ is boxed.  We call this {\em Case $1i$.}   There are three more subcases of  Case 1,
which we denote by {\em Case $1j$,  Case $1k$, and Case $1l .$ }
Case 1$i$, (1$j$, 1$k$,  or 1$l$,  respectively),
corresponds  to $\theta_{-i}$ ($\theta_{-j}$,  $\theta_{-k}$, or $\theta_{-l} $, respectively)
being so large that none of the terms in Table \ref{table.diffdelta1} involving it (i.e. all terms in one
column of the table) are a minimum in their row.    Due to the assumption  $a < b< c$, the choice of boxed terms
in Case 1$i$ is unique even for the first row of the table.   The main subcase and cases 1$i$ through 1$l$
comprise all possibilities.
\begin{table}[h]
  \caption{Case 1$i$ is indicated by boxes.   \label{table.diffdelta1i}   }
 $$
 \begin{array}{c|cccc}
 i  &                                   &	\theta_{-j} - a  - a                       & \framebox{$\theta_{-k}  - a  -a$} &  \framebox{$ \theta_{-l}  - a -a   $}   \\
j  &    \theta_{-i}   - b -b  &	                                                     & \framebox{$\theta_{-k}  - a  - b$}  & \framebox{$ \theta_{-l}  - a - b   $}  \\
k &   \theta_{-i}   - b -c   & \framebox{$\theta_{-j} -a - c $}  &                                                           &    \framebox{$  \theta_{-l}   - a  - b $}  \\
l  &    \theta_{-i}  - b  - c & \framebox{$  \theta_{-j} -a  - c $}  &  \framebox{$ \theta_{-k} - a - b   $}  &
 \end{array}
 $$
 \end{table}
 Each of the five subcases of Case 1 corresponds to a maximal polyhedron contained in the set of
 all substitute valuations.   All five subcases of Case 1 can be combined into the following
 set of conditions: \\
 
  {\bf  Case 1:}
 \parbox[t]{5in}{
 \begin{description}
 \item $
a-b \geq \min\{\theta_{-k}, \theta_{-l} \} - \min\{ \theta_{-i} , \theta_{-j}+b-a \} \geq a-c
 $, where
 \item  either the first inequality holds with equality, or $\theta_{-k} = \theta_{-l}$, and
 \item  either the second inequality holds with equality, or $\theta_{-i} = \theta_{-j}+b-a$.
 \item $\min\{\theta_{-k}, \theta_{-l} \} \geq a+b $  (needed for submodularity)
 \item $\min\{\theta_{-i}, \theta_{-j} + b-a \} \geq b+c $  (needed for submodularity)
 \end{description}
 }

For Case 2 we assume that $a<\min\{b,c\}$, for, as can be checked at the end, if this assumption
does not hold, then $v$ is in multiple maximal polyhedrons. 
With this asumption, Case 2 can similarly be
divided into five subcases, with each subcase corresponding to a maximal
 polyhedron in ${\cal S}_4$.   In the main subcase of Case 2, the boxed items in Table
 \ref{table.diffdelta2} are the minimums in their rows.
  \begin{table}[h]
  \caption{Specialization of Table \ref{table.diffdelta}   to Case 2, with the main subcase indicated by boxes.     \label{table.diffdelta2} }
 $$
 \begin{array}{c|cccc}
 i  &       &	\framebox{ $ \theta_{-j} - b  - a$}   & \theta_{-k}   - a  -a &  \framebox{$ \theta_{-l}  - a -b$}   \\
j  &   \framebox{$ \theta_{-i}   - a -c  $} &	    &\framebox{$ \theta_{-k}  - a  - c$}  & \theta_{-l}  - a - a  \\
k &    \theta_{-i}   - a -a   &   \framebox{$\theta_{-j} -b - a  $}  &      &  \framebox{$ \theta_{-l}   - b - a $}  \\
l  &    \framebox{$\theta_{-i}  - c - a $}   &	  \theta_{-j} -a  - a  & \framebox{$ \theta_{-k} - a - c  $}   &    
 \end{array}
 $$
 \end{table}
All five subcases of Case 2 can be combined into the following:\\
{\bf Case 2:}
 \parbox[t]{5in}{
 \begin{description}
 \item $
 b-a \geq \min\{\theta_{-j}, \theta_{-l} \} - \min\{ \theta_{-i} , \theta_{-k} \} \geq a-c,$ where
 \item  either the first inequality holds with equality, or $\theta_{-j} = \theta_{-l}$, and
 \item  either the second inequality holds with equality, or $\theta_{-i} = \theta_{-k}$.
 \item $\min\{\theta_{-j}, \theta_{-l} \} \geq a+b $  (needed for submodularity)
 \item $\min\{\theta_{-i}, \theta_{-k} \} \geq a+c $  (needed for submodularity)
 \end{description}
 }
 
 \begin{prop}
 There are 75 maximal polyhedrons comprising  ${\cal S}_4$, and each is
 ten-dimensional.    Of the 75 maximal polyhedrons, there are 12 corresponding
 to each of the five subcases of Case 1, and three corresponding to each of the
 five subcases of Case 2.
 \end{prop}
  \begin{proof}
 If Case 1 holds and $a < b < c$, then vertices $i$ and $j$ are uniquely identified, but
if vertices $k$ and $l$ are swapped, Case 1 still holds.   Consequently, there are
12 ways that Case 1 can hold, due to four possibilities for which vertex to label
$i$, and then three possibilities  of which vertex to label $j$,
and, as described above, five subcases for each labeling of the vertices.   
 If $a < \min\{b,c\}$, then up to
exchanging values of $b$ and $c$, there are three ways that Case 2 can hold,
corresponding to the three ways to select the pair of non-overlapping edges having
values other than $a$.  And there are five subcases of Case 2 for each possibility.
The maximal polyhedrons comprising ${\cal S}_4$ are constrained by three equalities
among the $\delta_{ij}$'s, two equalities among the $\theta_{-i}$'s, and
the equality $v(\emptyset)=0$, leaving ten out of 16 dimensions remaining.
\end{proof}
 
 \begin{rem}
The above parameterization suggests a simple way to generate substitute valuations on four
goods.    First choose $i,j,k,l$ to be a random permutation of $1,2,3,4$.    Then
select which of the ten subcases should hold, and then generate an element
of the corresponding polyhedron.
\end{rem}

\section{Generation of substitute valuations by Monte Carlo simulation}  \label{sec.MonteCarlo}
 
 An algorithm for generating a nondecreasing substitute valuation on a computer using a
random number generator is presented in this section.
The algorithm is formulated using the representation
 $v(A)=\mu\cdot A -\theta(A)$, in terms of $v$'s interaction function $\theta$ and vector of valuations of
 singleton sets, $\mu$.    
That is, the algorithm produces $(\theta(A): A\in 2^{\cal K} )$ with $\theta(A)=0$ for $|A|\leq 1$
so that for all bundles $A$ and distinct goods $i,j$ not in $A$:
\begin{equation}   \label{eq.submod} 
\theta(Aij)~ \geq~ \theta(Ai) + \theta(Aj) -\theta(A)
\end{equation}
and for all bundles $A$ and distinct goods $i,j,k$ not in $A$:
\begin{equation}   \label{eq.S3}
\theta(Aij)+\theta  (Ak)~ \geq ~ \min \{ \theta(Aik)+\theta(Aj), \theta(Ajk)+\theta(Ai) \},
\end{equation}
and it produces $\mu$ so that for all goods $k$:
\begin{equation}    \label{eq.mu}
\mu(k) \geq \max_{A:k\not\in A} \theta(Ak)-\theta(A).
\end{equation}


\begin{algorithm}  \caption{Substitute Valuation Generation Algorithm} \label{algorithm_one}
\fbox{
\parbox{5.5in}{
{\bf Step 1} Generate $\theta_o$ and $\mu_o$.  Initialize $\theta$ to $\theta_o.$\\
{\bf Step 2} for   $2\leq L \leq K$   \\
 $~~~~~~~~$for bundles $A$ with $|A|=L-2$ and distinct $i,j\not\in A$  \\
  $~~~~~~~~~~~~~~$increase $\theta(Aij)$ by the minimum amount so \eqref{eq.submod}  holds. \\
 $~~~~~~~~$while changes, for bundles $A$ with $|A|=L-2$ and distinct $i,j,k\not\in A$  \\
  $~~~~~~~~~~~~~~$increase $\theta(Aij)$ by the minimum amount so \eqref{eq.S3}  holds. \\
{\bf Step 3}  for $1\leq k \leq K$  \\
 $~~~~~~~~\mu(k) := \max\{ \mu_o(k), \max_{A:k\not\in A} \theta(A)-\theta(Ak)\}$.\\
}
}
 \end{algorithm}

The algorithm is shown in brief in the box, and is now explained in more detail.
In Step 1, nominal values $\theta_o$ and $\mu_o$ are generated. 
To avoid problems with roundoff error and to insure convergence, we require
$\theta_o$ and $\mu_o$ to be integer valued.  The $\theta$, $\mu$, and $v$ produced by
the algorithm will also be integer valued.
 The nominal valuation $v_o = \mu_o \cdot A -\theta_o(A)$ need not have the substitute
 property.     For example, we could take the value $\theta_o(A)$ for each bundle $A$
 to be a random variable  which is uniformly distributed over the interval of integers $[0, m|A| ]$
 for some constant $m$, or the sum of $|A|$ independent random variables, each uniformly
 distributed over the  interval $[0,m]$ for some $m$.  These suggestions are rather arbitrary,
 but they give larger means and variances for larger bundles.  In a particular application, there may
 be {\em a priori} knowledge about the typical distribution of the valuations which
 could be incorporated into this step.  The nominal values $\theta_o(A)$ need not be independent.

Step 2 of the algorithm produces $\theta$ in $K-1$ phases, indexed by $L$ running from 2 to $K$, and
each phase has two parts.
In  the first part of phase $L$, for each bundle $A$ with
$|A|=L-2$, and each choice of distinct goods $i$ and $j$ not in $A$, the following statement
is executed:
\begin{equation}  \label{eq.super}
\theta(Aij) :=  \max \{ \theta(Aij), \theta(Ai) + \theta(Aj) - \theta(A)  \}.
\end{equation}
This insures supermodularity of $\theta$ up to level $L$.
There is no need to visit a particular $A$, $i$, $j$ more than once in this
part of the phase. 
The second part of phase $L$ consists of one or more iterations.   In each iteration,
for each bundle $A$ with $|A|=L-2$, and each choice of distinct goods $i,j,k$
not in $A$, the following statement is executed:
\begin{equation}  \label{eq.S_iter}
\theta(Aij) :=   \max \left\{\theta(Aij) ,  \min\{ \theta(Aik) + \theta(Aj), \theta(Ajk) + \theta(Ai) \}- \theta(Ak) \right\}.
\end{equation}
Multiple iterations may be needed in the second part of phase $L$ because
of the terms  $\theta(Aik)$ and $\theta(Ajk)$, which also involve $\theta$ evaluated on sets
of cardinality $L$, in the righthand side  of \eqref{eq.S_iter}.   These terms may be
increased after the statement for $\theta(Aij)$ is executed, during the same iteration, which
can require that $\theta(Aij)$ be increased again later. The inclusion of the term $\theta(Aij)$
on the righthand sides of \eqref{eq.super} and \eqref{eq.S_iter} insures that the $\theta$ values are nondecreasing
during execution of the algorithm.  If there are no
changes during an iteration, then the second part of phase $L$ of the algorithm is complete. 
This happens if and only if $\theta$ satisfies condition $S3_{\theta}(L)$ before the iteration.
By the nature of the algorithm,  the proof  of correctness comes down to a proof that the
number of iterations needed in each phase of Step 2 is finite.  

Step 3 of the algorithm sets $\mu$ to the smallest vector greater than or equal to $\mu_o$ such
that  $ \eqref{eq.mu}$  is satisfied.  The description of the algorithm is complete.

Note that randomization is used only in Step 1 of the algorithm.   The deterministic portion of the
algorithm, Steps 2 and 3, insure that the $\theta$ and $\mu$ produced satisfy the substitutes and
monotonicity conditions.   There may be other applications for the deterministic portion of the algorithm.  
For example, the true valuation of a buyer may not quite be a substitute valuation, and participation
in a particular auction may require an input that is a substitute valuation.  Then the deterministic
portion of the algorithm could be run to find a substitute valuation close to the original valuation.

We define a partial ordering ``$\prec$" on the set of interaction functions,
which is pointwise within levels and lexicographic among levels, as follows.
Write $\theta'  \prec \theta $ if $\theta'=\theta$ or if there exists $L$ with $2\leq L \leq K$ such
that $\theta(A)=\theta'(A)$ if $|A|<L$,  $\theta(A) \leq \theta'(A)$ if $|A|=L$, and
 $\theta(A_o) <  \theta'(A_o)$ for some $A_o$ with $|A_o|=L$.
 
\begin{prop}  \label{prop.alg_correct}
The algorithm terminates in finite time, and the corresponding valuation $v(A)=\mu\cdot A - \theta(A)$ is
a nondecreasing substitute valuation.    The interaction function $\theta$ produced is  minimal
in the  ``$\prec$" order,  among all supermodular interaction functions satisfying $S3_\theta$ which
pointwise dominate  $\theta_o$.
\end{prop}

A proof of Proposition \ref{prop.alg_correct} is given in this section, but first some
properties connected with substitute valuations are given.

\begin{defn} (Property $F4_{\theta}(L)$)  Let $2 \leq L \leq K-2$.   An interaction  function $\theta$ is said
to have property $F4_{\theta}(L)$ if
$(\theta(Aij)+\theta(Akl) , \theta(Aik) + \theta(Ajl), \theta(Ail)+\theta(Ajk) )$
has the double minimum property
whenever $A$ is a bundle with $|A|=L-2$, and $i,j,k,l$ are distinct goods not in $A$.
\end{defn}

\begin{lem}   \label{lemma.S3_to_F4}
Let $2 \leq L \leq K-2$.  If $\theta$ satisfies $S3_{\theta}(L)$ then it satisfies $F4_{\theta}(L).$
\end{lem}
\begin{proof}
This result for $K=4$ was already stated as Lemma \ref{lemma.F4}.
That is,  if ${\cal K}=\{i,j,k,l\}$,
then the fact that the double minimum property is satisfied by any three of the
six numbers $\delta_{ij}, \delta_{ik}, \delta_{il}, \delta_{jk}, \delta_{jl}, \delta_{kl}$
corresponding to a triangle, implies the double minimum property for the triple
$( \delta_{ij}+  \delta_{kl}, \delta_{ik}+ \delta_{jl}, \delta_{il} + \delta_{jk}).$
In general, for $2 \leq L \leq K-2$, for a bundle $A$ fixed with $|A|=L-2$, and for distinct goods $i,j,k,l$
not in $A$, condition $S3_{\theta}(L)$, which is equivalent to condition $S3_{\delta}(L)$,
implies that the double minimum property is satisfied by any three of the
six numbers $\delta_{ij|A}, \delta_{ik|A}, \delta_{il|A}, \delta_{jk|A}, \delta_{jl|A}, \delta_{kl|A}$
corresponding to a triangle.   So by the same reasoning used for Lemma \ref{lemma.F4}, $S3_{\theta}(L)$
implies that
$( \delta_{ij|A}+  \delta_{kl|A}, \delta_{ik|A}+ \delta_{jl|A}, \delta_{il|A} + \delta_{jk|A})$
satisfies the double minimum property, and thus that $F4_{\theta}(L)$ holds.
\end{proof}

\begin{lem}   \label{lemma.F4_to_S3}
Suppose either $L=1$,  or $2 \leq L \leq K-1$
and $(\theta(A): |A| = L)$ satisfies $F4_{\theta}(L).$
Also, suppose $\mu \in \IR^{\cal K}$, 
and suppose $\theta$ is determined on sets $A$ with $|A|=L+1$ as follows:
\begin{equation}  \label{eq.extension}
\theta(A) = \min_{i\in A}~ ( \theta(A-i) + \mu(i) ).
\end{equation}
Then $\theta$ satisfies $S3_{\theta}(L+1).$
\end{lem}

\begin{proof}
Let $|A|=L-1$ and suppose $i,j,k$ are distinct goods not in $A$.   Let $l^*$ be a
good in $Aij$ such that $\theta(Aij)=\theta(Aij-l^*)+\mu(l^*)$.  It must be shown that
\begin{eqnarray}  \label{eq.S3F4}
\theta(Aij-l^*) + \mu(l^*)  +\theta(Ak)  \geq  ~~~~~~~~~~~~~~~~~~~~~~~~~~~~~~~~~~~~~~~~~~~~~~~~~~~~~~~~~ \\
\min\left\{
\min_{l'\in Aik} \theta(Aik-l') + \mu(l') + \theta(Aj)   , \min_{l'' \in Aik} \theta(Ajk-l'') + \mu(l'') + \theta(Ai)  \right\}.
\nonumber
\end{eqnarray}
If $l^*=i$, then trivially,
$$
\theta(Aij-l^*) + \mu(l^*) + \theta(Ak)  = \theta(Aik-l^*) +  \mu(l^*) + \theta(Aj)   .
$$
Since $l^*$ is a possible value of $l'$ in \eqref{eq.S3F4},  \eqref{eq.S3F4} follows.
Similarly,  \eqref{eq.S3F4}  is true if $l^*=j$.   So for the remainder of this proof we suppose
that $l^* \in A$, which can happen only if $L\geq 2$.  Let $B=A-l^*$.   By $F4_{\theta}(L)$,
$$
\theta(Bij)+\theta(Bl^*k) \geq \min \{ \theta(Bik)+ \theta(Bjl^*), \theta(Bjk)+\theta(Bil^*)\}
$$
which is equivalent to
\begin{eqnarray*}
\theta(Aij-l^*)+\mu(l^*)  +\theta(Ak)  \geq  ~~~~~~~~~~~~~~~~~~~~~~~~~~~~~~~~~~~~~~~~~~~~~~~~~~~~~~~~~ \\
\min \{ \theta(Aik-l^*)+\mu(l^*)  + \theta(Aj)   , \theta(Ajk-l^*) +\mu(l^*)  +\theta(Ai)   \}.
\end{eqnarray*}
Since $l^*$ is a possible value for either $l'$ or $l''$, this implies \eqref{eq.S3F4}.
\end{proof}

\begin{proofof}  {\bf Proposition \ref{prop.alg_correct}. }
Step 3 insures that the output valuation is nondecreasing, so it is sufficient to show that Step 2
of the algorithm eventually terminates, and the interaction function $\theta$ produced satisfies
the properties advertised in the proposition.   Suppose $\theta'$ is any supermodular interaction
function satisfying $S3_{\theta}$ which pointwise dominates $\theta_o$.
During phase $L$, the algorithm only modifies values of $\theta(B)$ with $|B|=L$, and it does
so by increasing the values the smallest amount possible, so as to satisfy the supermodularity
and $S3_{\theta}(L)$ conditions.   If phase $L$ is completed,  $\theta$ must be supermodular
up to level $L$ and satisfy $S3_{\theta}(L)$.   Moreover, if $\theta(B)=\theta'(B)$
for all sets $B$ with $|B|<L$, then $\theta(B) \leq \theta'(B)$ for all sets $B$ with $|B|=L$.
Therefore, if the algorithm terminates, the interaction function $\theta$ produced must have
the desired properties. 

It remains to show that the algorithm terminates.  To that end,  it will be shown by induction on
$L$ that, for $1 \leq L \leq K$, the following statement
is true:  Either $L=1$ or the algorithm completes phase $L$.
The statement is trivially true for the base case $L=1$.  For the sake of argument by induction,
suppose the statement is true for some $L$ with $1 \leq L \leq K-1$.    If $L\geq 2$, then
at  the end of execution of phase $L$,  $\theta$ satisfies $S3_{\theta}(L)$, and
hence also $F4_\theta(L)$,  by Lemma \ref{lemma.S3_to_F4}.   Thus, either $L=1$,
or $2\leq L \leq K-1$ and $\theta$ satisfies $F4_{\theta}(L)$.   Let 
$$
\overline{\theta}(A)=\left\{ \begin{array}{cl} \theta(A) & \mbox{if}~|A|\leq L \\
                  c+\min_{i\in A} \theta(A-i) & \mbox{if}~|A| = L+1,
                    \end{array}  \right.
$$ 
where $c$ is a constant chosen large enough that $\overline{\theta}$ is
supermodular up to level $L+1$, and
$\overline{\theta}(A) \geq  \theta_o(A)$ for all $A$ with $|A|=L+1$.  By Lemma 
\ref{lemma.F4_to_S3} with $\mu(i) = c $ for all $i$,  it follows
that $\overline{\theta}$  satisfies $S3_{\theta}(L+1).$   It follows that $(\theta(A): |A|=L+1)$
is bounded above by $(\overline{\theta}(A) : |A|=L+1)$ throughout execution
of phase $L+1$ of the algorithm.   Since some entry in $(\theta(A): |A|=L+1)$ strictly increases
each time the algorithm finds a violation of $S3_{\theta}$, since $(\theta(A) : |A|=L+1)$
is bounded above, and since all the values are integers, execution of phase  $L+1$ must terminate in a
finite number of steps.
Therefore, the induction statement is true for $L+1$, and hence for all $L$ in the
range $1\leq L \leq K$, as required.
\end{proofof}

\begin{rem}   \label{remark.large_iter}
The number of iterations required by the algorithm was proved to be finite by
appealing to the assumption that integer values are used.
Unfortunately, the number of iterations of the algorithm is not bounded by a function of
$K$ alone.    Indeed, if $K=6$, and the nonzero values of $\theta_o$
are  $\theta_o(111100)=\theta_o(110011)=\theta_o(001111)=m\geq 1$ and
$\theta_o(000111)=\theta_o(010111)=\theta_o(100111)=1$, then $\theta_o$
satisfies $S3_{\theta}(2)$ and $S3_{\theta}(3)$ and $\theta_o$ is submodular
up to level $L=4$.   Thus, at the beginning of the second half of phase $L=4$ of Step 2,
$\theta=\theta_o$.   Starting from that point, for  $1\leq j \leq m$, after $j$ iterations,
$\theta(A)=j$ for all $A$ with $|A|=4$ and $A\not\in  \{111100,110011,001111\},$
and $\theta$ is still equal to $m$ on $ \{111100,110011,001111\}$.   After $m$ iterations,
$\theta(A)=m$ for all $A$ with $|A|=m$, and no changes are made during the $(m+1)^{th}$
iteration.
Therefore, $m+1$ iterations are needed.  Since $m$ is an arbitrary positive integer, the number
of iterations is thus not bounded by a function of $K$ alone.
However, for all distributions of $\theta_o$ that we tried along the lines of those we suggested for
Step 1, the average number of iterations per level was less than 2.1 per level,  for up to $K=20$.   
It was only through running millions of examples that we discovered the behavior exhibited in this example.
\end{rem}  

\begin{rem}  \label{remark.Mconcave}
Condition $F4_{\theta}(L)$ for $L$ fixed, when translated to a condition on $v$,
 is a special case of the  local exchange property $EXC_{loc}$ introduced by
 Murota  \cite[p. 282]{Murota96se}.
As noted at the end of Section \ref{sec:background}, substitute valuations are $M^\natural$ concave
functions.   It follows that valuations restricted to a single level $\{ A \in 2^{\cal K} : |A|=L\}$
are $M$-concave (in fact they are valuated matroids).    Murota \cite{Murota96se} showed
that the local exchange property implies the seemingly more general exchange
property in the definition of $M$-concavity.    
\end{rem}

As a by-product of the proof of convergence of Proposition \ref{prop.alg_correct}, we recover the
following result about $L$-satiation (defined in Section \ref{sec:background}):
\begin{cor}  \label{cor.LSAT}  \cite{BingLehmannMilgrom04}
Let $0\leq L \leq K$.  Then the $L$-satiation of a substitute valuation is also
a substitute valuation.
\end{cor}
\begin{proof}
Let $\widehat{v}$ be a substitute valuation, and let $v$ denote its $L$-satiation.
To avoid trivialities, assume $1\leq L \leq K-1$.
Since $v$ equals $\widehat{v}$ up to level $L$,
$v$ is nondecreasing, is submodular, and satisfies $S3$, all up to level $L$.
That is,
\begin{enumerate}
\item$v(A) \leq v(Ai)$ whenever $A$ is a bundle with $|A| \leq L-1$ and $i$ is a good not in $A$
\item
$v(Aij)-v(Ai)-v(Aj)+v(A) \leq 0$ whenever $A$ is a bundle with $|A| \leq L-2$ and $i$ and $j$
are distinct goods not in $A$
\item  $S3(L')$ holds for $0\leq L' \leq L$.
\end{enumerate}
It suffices to show that $v$ is nondecreasing, submodular, and satisfies $S3$, all up to level $L+1$, for then an obvious proof by induction can be used to show that these properties hold
for all levels, implying that $v$ is a substitute valuation.

Clearly,  $v$ is nondecreasing up to level $L+1$.
To prove that $v$ is submodular up to level $L+1$, let $A$ be any bundle with $|A|=L-1$,
and let $i$ and $j$ be goods not in $A$.  It must be shown that for any item $k \in Aij$,
$v(Aij-k)-v(Ai)-v(Aj)+v(A) \leq 0$.   If $k \in \{i,j\}$ this inequality is true because $v$ is
nondecreasing.   If $k\in A$, then with $B=A-k$, the inequality reduces to
\begin{equation}  \label{eq.Bineq}
v(Bij)-v(Bik)-v(Bjk)+v(Bk) \leq 0
\end{equation}   
But, by property $S3(L)$,  either
 $v(Bij)+v(Bk) \leq v(Bik)+v(Bj)$ or  $v(Bij)+v(Bk) \leq  v(Bjk)+v(Bi)$.  Either one
 of these inequalities and the fact $v$ is nondecreasing implies \eqref{eq.Bineq}.
 Therefore, $v$ is submodular up to level $L+1$.

By Lemma \ref{lemma.S3_to_F4}, the interaction function of $v$ satisfies $F4_{\theta}(L)$.
Let the vector $\mu$ in the construction  \eqref{eq.extension}
of Lemma \ref{lemma.F4_to_S3} denote the single item price vector for $v$.
Then, with $\theta$ denoting the interaction function of $v$,   \eqref{eq.extension}
is equivalent to $v(A)=\max_{i\in A} v(A-i) $.   That is, the extension formula
\eqref{eq.extension} is the same as the definition of $v$ on level $L+1$ as the $L$-satiation of
$\widehat{v}$.   Therefore, Lemma \ref{lemma.F4_to_S3} implies that $v$ has property
$S3(L+1)$. 
\end{proof}

\begin{rem}  Submodularity by itself is not necessarily preserved by $L$-satiation.
For example, the following valuation is nondecreasing and submodular (but is not a substitute
valuation): 
$$
\begin{array}{c|cccccccccccccccc}
A & \emptyset & i & j & k  & l & ij & ik & il & jk & jl & kl & ijk & ijl & ikl& jkl  & ijkl\\  \hline
\widehat{v}(A)   &  0          & 3 & 3 & 3  &  3  & 4 & 4  & 4 & 4 & 4 & 5 & 5 & 5 & 5 & 5 & 5
 \end{array}
 $$
Its 2-satiation satisfies $v(ijkl)-v(ijk)-v(ijl)+v(ij) > 0$, and so it  is not submodular.
\end{rem}

\section{Assignment valuations}    \label{sec.assign_val}

Assignment valuations form an important subclass of substitute valuations, and they
can be described as follows.
An economy of $n$ buyers, each with a single unit valuation, can be represented by an
$n\times K$ weight matrix $W$, with entry $w_{i,k}$ denoting the value of good $k$ to buyer $i$.
The aggregate valuation $v=v_1 *  \cdots *v_n$
for such a set of buyers is determined by an assignment problem for each bundle  $A$, described
as follows.  For notational convenience, let $\triangle$ be a null item, not in $\cal K$,  and let
$w_{i \triangle}=0$ for any buyer $i$.
For any $A \subset  {\cal K}$, an assignment of the goods in $A$ to the $n$ buyers is given
by a mapping $\sigma : \{1, \cdots , n \} \rightarrow A \cup \{ \triangle \}$ such that
$\sigma_i = \sigma_j$ only if $\sigma_i = \sigma_j=\triangle$.  The interpretation is that buyer
$i$ is allocated good $\sigma_i.$   All goods in $A$ need not be allocated.   Let $T(A)$ denote
the set of such assignments.  Then  $v(A) = \max \{  \sum_{i=1}^n  w_{i,\sigma_i} : \sigma \in T(A) \}$.
A valuation $v$ arising from a weight matrix $W$ in this way is called an
{\em assignment valuation}.    Assignment valuations are an important subclass of
substitute valuations.  Special cases of assignment valuations include the linear valuations and the separable concave valuations, mentioned in Section \ref{sec.introduction}.

\subsection{Assignment valuations with monotone assignments}   \label{sec.mono_assign}

 If the weight matrix $W$ for an assignment valuation $v$ satisfies certain conditions, then
there is a very simple formula for the valuation.  Specifically, suppose $W$ has dimension
$K \times K$ with the following properties:
 \begin{enumerate}
  \item (Nonnegative)  $w(i,k)\geq 0$   for $1\leq i \leq K$, $1\leq k \leq K,$
 \item (Nonincreasing in $i$)  $w(i+1,k) \leq w(i,k)$  for $1\leq k \leq K$, $1\leq i \leq K-1,$
  \item  (Supermodular)   $w(i+1,k+1)-w(i,k+1)-w(i+1,k)+w(i,k) \geq 0$
  for $1\leq i \leq K-1$, $1\leq k \leq K-1.$

\end{enumerate}
An example for $K=5$ is as follows:
$$
W=\left( \begin{array}{ccccccc}

32 & 35 & 25 & 26 & 22  \\
24 & 30 & 20 & 21 & 19 \\
16 & 22 & 14 & 16 & 14  \\
9 & 15 & 7 & 9 & 9   \\
2 & 8 & 1 & 4 & 5   \\
  \end{array}
  \right) .
$$
 Writing a bundle $A$ as  $A=\{k_1, \ldots, k_L \} $ with $k_1 < \cdots  < k_L $, we claim
  that an optimal assignment for $A$ is given by $\sigma_i = k_i$ for
 $1\leq i \leq L$.   That is, for $1 \leq i \leq L$, the $i^{th}$ buyer is allocated the $i^{th}$
 good in $A$.    To see this, note by the monotonicity and nonnegativity that the goods
 should be assigned to buyers 1 through $L$, and then the supermodularity implies
 that if $1 \leq i < i' \leq L$  and if buyer $i$ is assigned a good with a higher index than
 the good assigned  to buyer $i'$, then the value of the assignment is not decreased if the
 goods are swapped.  
 
 Hence, the valuation $v$ can be expressed in terms of $W$ as follows:
$$
v(\{k_1,\cdots , k_L\} ) =   \sum_{i=1}^{L} w(i,k_i) .
$$
An interpretation is that there is a variable value for each item $k$ included in the assigned set.
 The variable value for an item $k$ is
$w(i,k)$ if $k$ is the $i^{th}$ good in the bundle.
 
Since the maximum matchings do not involve $w(i,k)$ for $k<i$, the same matchings are
still maximum weight matchings with the same values if
$W$ is changed by setting $w(i,k)=0$ for $k<i$, to obtain a matrix of the form:
$$
\widehat{W}=\left( \begin{array}{ccccccc}

32 & 35 & 25 & 26 & 22  \\
0 & 30  & 20 & 21 & 19 \\
0 & 0  & 14 & 16 & 14  \\
0 & 0 & 0 & 9 & 9   \\
0 & 0 & 0 & 0  & 5   \\
  \end{array}
  \right) .
$$
The matrix $\widehat{W}$ satisfies the following conditions:
 \begin{description}
 \item ($\widehat{0}$) (Upper triangular) $\widehat{w}(i,k)=0$ for $1\leq k < i \leq K,$ 
\item  ($\widehat{1}$) (Nonnegative)  $\widehat{w}(i,k)\geq 0$   for $1\leq i\leq K$, $1\leq k \leq K,$
 \item  ($\widehat{2}$) (Supermodular on upper triangle)   $\widehat{w}(i+1,k+1)-\widehat{w}(i,k+1)-\widehat{w}(i+1,k)+\widehat{w}(i,k) \geq 0$
  for $1 \leq i \leq K-1$, $i < k  \leq K-1,$
 \item ($\widehat{3}$) (Nonincreasing in $i$ for $k=K$)  $\widehat{w}(i+1,K) \leq \widehat{w}(i,K)$  for $1\leq i \leq K-1,$
 \item ($\widehat{4}$)  (Extension condition)  $(\widehat{w}(k,k)+ \widehat{w}(k+1,k+1) \cdots + \widehat{w}(K,K) ) - \\(\widehat{w}(k,k+1)+\widehat{w}(k+1,k+2)+\cdots +\widehat{w}(K-1,K)) \geq 0$ for $1\leq k \leq K-1.$
\end{description}
Moreover, in general, the reverse direction can be taken:
\begin{prop}  \label{prop.W_hat}
Suppose a weight matrix $\widehat{W}$ satisfies the conditions $\widehat{0}-\widehat{4}$ above.
Then an optimal assignment for any bundle  $A=\{k_1, \ldots, k_L \} $
 with $k_1 < \cdots  < k_L $, is given by $\sigma_i = k_i$ for  $1\leq i \leq L$.  
 \end{prop}

\begin{proof}
It suffices to show that $\widehat{W}$ can be modified in positions  $i > k$ so that the modification
$W$ satisfies the original conditions 1-3.   
Working from right to left, it is clear that if $\widehat{W}$ is to be
modified for indices $i>k$ so that the modification is supermodular everywhere, then the
maximal such modification is such that  $\widehat{w}(i+1,k+1)-\widehat{w}(i,k+1)-\widehat{w}(i+1,k)+\widehat{w}(i,k) = 0$ for $1\leq k < i \leq K.$
Such a supermodular modification of $\widehat{W}$ will be nonincreasing in $i$ for all $k$ because
of the supermodularity and condition $\widehat{3}$.   It remains to check that the modification is
nonnegative.   However, the quantities in condition $\widehat{4}$ are the first $K-1$ entries of the
last row of the modified matrix.
Thus, the modification $W$ of $\widehat{W}$ satisfies conditions 1-3, as claimed.
\end{proof}

See \cite[Section 3.2]{Topkis98} for much more general versions of this monotonicity result.

Just as the set of all substitute valuations ${\cal S}_K$ can be represented
as a union of finitely many polyhedrons, the same is true for the set of assignment valuations.
Indeed, there is a finite number of ways to select an assignment $\sigma$ for each set
of goods $A$.   Some selections are the optimal ones for a nonempty set of weight matrices
$W$, which forms a polyhedron within the set of weight matrices.   The corresponding
valuations for the fixed selections thus form a polyhedral subset of ${\cal S}_K$.
The union of such sets is precisely the set of assignment valuations.

There are $K(K+1)/2$ degrees of freedom in the choice of $\widehat{W}$ for the
weight matrices described in Proposition \ref{prop.W_hat}.   Thus, $K(K+1)/2$ is a lower bound
on the largest dimension of a polyhedral subset of the set of assignment valuations.
An upper bound is $K^2-(K-1)$, which can be seen as follows.  For each good, we can
form a list of the buyers according to decreasing weight, with ties broken arbitrarily. 
If a given good is assigned to some buyer, the other buyers higher on the list for that good should
also be assigned to some good.   It can be seen that at most one good would ever be forced to be
assigned to its $K^{th}$ choice of buyer.

\subsection{Assignment valuations for four goods}  \label{sec.four_assign}

The following proposition shows the relationship between the maximal polyhedrons
of the set of substitute valuations and the maximal polyhedrons of the set of assignment valuations,
for four goods.

\begin{prop}
All five subcases of Case 1 for $K=4$ correspond to assignment valuations, and in particular
the valuations of Case 1i are those that arise from Proposition \ref{prop.W_hat} for $K=4$,
but none of the valuations of Case 2 with $0 < a < \min\{b,c\}$ are assignment valuations.
That is, the maximal polyhedrons of assignment valuations in ${\cal S}_4$ are precisely
the 60 maximal  polyhedrons of ${\cal S}_4$ corresponding to Case 1.
\end{prop}

\begin{proof}
The valuations in the main subcase of Case 1 correspond to assignment valuations for the
ordering of states $(i,j,k,l)=(1,2,3,4)$ and weight matrices of the form
$$
W=\left( \begin{array}{llll}
\mu_1 & \mu_2 & \mu_3 & \mu_4  \\
\mu_1-a & \mu_2-b & \mu_3-c  & 0 \\
0 & \mu_2-d & \mu_3-e & 0 \\
0 & 0 & \mu_3-f & 0
 \end{array}
 \right).
 $$
 under the following constraints on the constants involved.
The constants $a,b,c$ and the vector $\mu$ have the same significance as in Case 1.
The constraints  on  $a,b,c,d,e,f$ are $0 \leq a \leq b \leq c \leq  e \leq f$ and $0 \leq e-d \leq c-b,$
while the constants $d,e,f$ parameterize the remaining three degrees of freedom in the
choice of $\theta$.
The vector $\mu$ must be large enough so that  for any bundle $A$, it is optimal
to assign all goods in $A$ for the purposes of computing $v(A)$.   Equivalently,
if only such full matchings are considered, the resulting valuation should be
nondecreasing in $A$.   In this case, it means that $\mu$ should satisfy the constraints:
$$\mu_1 \geq a+d-b+f-e, ~ \mu_2 \geq d+f-e, ~  \mu_3\geq f,  ~ \mu_4\geq f$$
In the description of other cases below, the constraints on $\mu$ are determined
similarly, without comment.  Under these constraints, the assignment valuation falls into the main
subcase of Case 1 with  $\theta_{-l}=\theta_{-k}=a+d$, $\theta_{-j}=a+e$, $\theta_{-i}=b+e$,  and
$\theta(1234) =a+d+f$.   It is not difficult to show that all valuations in the main subcase of
Case 1 can be so obtained.
 
The same weight matrix $W$ covers Case $1j$ under the conditions
 $0\leq a \leq b \leq c,$   $e \leq f $, and $0\leq d-b\leq e-c$.
 
 The same weight matrix $W$ covers Case $1k$ under the conditions
 $0\leq a \leq b \leq c \leq e \leq \min\{ d, f \}$.    Since $k$ and $l$ play a symmetric
 role in Case 1, we can cover Case $1l$ by using $W$ with the third and fourth
 columns interchanged.
 
 Finally, to obtain Case $1i$ we can use the weight matrix
 $$
W'=\left( \begin{array}{llll}
\mu_1 & \mu_2 & \mu_3 & \mu_4  \\
\mu_1-a & \mu_2-b & \mu_3-c  & 0 \\
\mu_1-d &   \mu_2-e   &0 & 0 \\
\mu_1-f & 0 & 0 & 0
 \end{array}
 \right).
 $$
 with the conditions $a \leq b \leq c $,  ~ $e-b  \geq d-a \geq 0,$  and $f\geq d$.
The set of possible values of the matrix $W'$ as the ten constants vary
(including the constraints on the $\mu_k$'s)  corresponds to the set of matrices $\widehat{W}$
given by Proposition \ref{prop.W_hat}, except with the columns listed in reverse order.
That is, for $K=4$, Proposition \ref{prop.W_hat} refers to the valuations of Case 1i.
 
Lehmann et al.  \cite[Example 1]{LehmannLehmannNisan05} gives an example of a
substitute valuation for $K=4$ which is not an assignment valuation.  The example falls
into Case 2 with $a=1$ and $b=c=5.$   The following argument, also used in
\cite{LehmannLehmannNisan05}, shows that none of the valuations
of Case 2 with $0 < a < \min\{b,c\}$ are assignment valuations.
For the sake of argument by contradiction, suppose $v$ is a valuation in Case 2
with $0 < a < \min\{b,c\}$, and that $v$ is an assignment valuation for a weight matrix $W$. 
Since all the pairwise $\delta$'s are strictly positive, some row of $W$ must equal
$(v(i),v(j),v(k),v(l))$, and each of these entries is the maximum entry in its respective column.  
Since $\delta_{ik} > a$, any other entry of column $k$
must be strictly less than $v(k)-a$.   Similarly, since $\delta_{jl} > a$, any other entry of column $j$
must be strictly less than $v(j)-a$.  But then it is impossible that $\delta_{jk}=a$.  Therefore,
as claimed, none of the valuations of Case 2 with $0 < a < \min\{b,c\}$ are assignment valuations.
\end{proof}
 
\section{Speckled valuations}   \label{sec.spec_val}

It turns out that some substitute valuations are significantly different from assignment
valuations, and they cover polyhedrons with substantially larger dimension.    

\begin{prop}  \label{prop.expo_substitutes}
There is a polyhedron contained in ${\cal S}_{16}$ with dimension 2,727, and
a polyhedron contained in ${\cal S}_{24}$ with dimension 424,607.   For any $K\geq 2$,
there is a polyhedron contained in ${\cal S}_K$ with dimension at least
$2K-1 +  \frac{2^{K-1}-2}{K}$.
\end{prop}

Before getting to the proof of the proposition, we shall introduce some terminology from
the theory of binary codes.   A codeword of length $K$
is a binary sequence of length $K$, and the weight of a codeword is the number of one's in
the codeword.    A bundle $A$ naturally corresponds to the codeword with a one in the
$k^{th}$ position if and only if good $k$ is in $A$, for $1 \leq k \leq K$.    Let
$d_H(A,A')$ denote the {\em Hamming distance} between two sets: 
i.e. $d_H(A,A') = |A \backslash A'|+|A'  \backslash  A|.$

The proof of the proposition is based on the following construction, valid for $K \geq 2$.   
Let $\alpha_1,  \cdots , \alpha_K$ and $\beta_2, \beta_3 , \ldots \beta_K$ be
constants in the interval $[0,1]$.  Let $\beta_0=\beta_1=0$.
Let ${\cal C}$ denote a collection of bundles such that
\begin{enumerate}
\item  For all $A\in {\cal C}$,  $2\leq |A| \leq K-1$  and $|A|$ is even.
\item If $A, A' \in {\cal C}$ and $|A|=|A'|$ then $d_{H}(A,A') \geq 4$.
\end{enumerate}
Let $\gamma_A$ be an element
of $[0,1]$ for any $A \in {\cal C}$.
Define the interaction function $\theta$ by
$$
\theta(A )=  \beta_{|A|} +  I_{ \{ A \in {\cal C} \} }  \gamma_A  +   \phi(|A|)
$$
where $\phi(L)=(1.5)L(L-1)$,
and define the vector $\mu$ by  $\mu_k=(3K-1)+\alpha_k$.
Let $v$ be the corresponding valuation: $v(A)=\mu\cdot A - \theta(A)$.
We call the valuations of this form {\em speckled valuations}, thinking of the
many values of $\gamma_A$ for $A\in {\cal C}$ as specks, or small spots,
on the valuation.
\begin{lem}
The valuation $v$ is a nondecreasing substitute valuation.  For $\cal C$ fixed,
there are $2K- 1 + |{\cal C}|$ degrees of freedom in the choice of $v$.
\end{lem}

\begin{proof}
It suffices to show that $\theta$ has property $S3_{\theta}$, that
$\theta$ is submodular, and that $v$ is nondecreasing.
Property $S3_{\theta}$ amounts to showing that
$(\theta(Aij)+\theta(Ak) , \theta(Aik) + \theta(Aj), \theta(Ajk)+\theta(Ai) )$ has the
double minimum property whenever $A$ is a bundle
and $i,j,k$ are distinct goods not in $A$.
For fixed $A,i,j,k$ with $|A|=L-2$,
this condition involves $\theta$ evaluated on three sets of cardinality $L-1$ and three
sets of cardinality $L$.     Moreover, the three sets of cardinality $L-1$ each have
Hamming distance two to the other two sets of cardinality $L-1$.   Likewise,
the three sets of cardinality $L$ each have
Hamming distance two to the other two sets of cardinality $L$.   Therefore, at most
one of the six sets involved is in $\cal C$. 
If none of the six sets is in $\cal C$, then
the three values $\theta(Aij)+\theta(k), \theta(Aik)+\theta(j), \theta(Ajk)+\theta(Ai)$
are equal.
  If one of the six sets is in $\cal C$, then 
$(\theta(Aij)+\theta(k), \theta(Aik)+\theta(j), \theta(Ajk)+\theta(Ai))$ still has
the double minimum property.  So $\theta$ has property $S3_{\theta}$.

To see that $\theta$ is supermodular, let $2 \leq L \leq K$,  let $A$ be a bundle with
cardinality  $L-2$, and let $i, j$ be goods not in $A$.  Note that $\phi(L)-2\phi(L-1)+\phi(L-2)=3$,
and also that at most one of $Ai$ and $Aj$ are in $\cal C$.  These
observations and the fact that the $\beta$'s and $\gamma$'s are in the interval
$[0,1]$ imply that 
$\theta(Aij)-\theta(Ai)-\theta(Aj)+\theta(A)   \geq  3  - 2\beta_{L-1} - 1 \geq 0$,
so that $\theta$ is supermodular.

As for the monotonicity of $v$, note that for $1\leq L \leq K$,  $\phi(L)-\phi(L-1)=3(L-1)\leq 3(K-1).$
So if $A$ is a bundle with some cardinality
$L-1$, and $i$ is a good not in $A$, then
$$v(Ai)-v(A) \geq 3K-1  - \beta_{L-1} - I_{\{ A \in {\cal C} \} }\gamma_A - 3(K-1) \geq 0.$$
Thus $v$ is nondecreasing.

For fixed $\cal C$, there are $2K- 1 + |{\cal C}|$ degrees of freedom in the
choice of the $\alpha$'s, $\beta$'s and $\gamma$'s.   It is easy to check that the
mapping from these variables to $v$ is linear and invertible.
\end{proof}

\begin{proofof}  {\bf Proposition \ref{prop.expo_substitutes}.}
For a fixed $\cal C$, the dimension of the set of valuations constructed above
is $2K-1 + |\cal C|$, so it remains to show that $|\cal C|$ can be taken large enough. 
The maximum possible cardinality of $\cal C$ subject to the
above conditions can be expressed as follows:
\begin{equation}  \label{eq.Cbnd}
| {\cal C} |  =  \sum_{L: L~\mbox{even},~2\leq L \leq K-1}  A(K,4,L),
\end{equation}
where $A(K,4,L)$ denotes the maximum possible cardinality of a set of
weight $L$ binary codewords of length $K$ with Hamming distance at least
4 between any two codewords.  By symmetry, $A(K,4,L)=A(K,4,K-L).$
Tables in \cite{BrouwerShearerSloaneSmith90} show that
$A(16,4,2)=8$, $A(16,4,4) \geq 140$,  $A(16,4,6) \geq 615$, 
and $A(16,4,8)\geq 1170$, so for $K=16$ it is possible that
$|{\cal C}| = 2(8+140+615)+1170 =2696$, giving the bound for $K=16$
in Proposition \ref{prop.expo_substitutes}.
Similarly, for $K=24$, it is possible that
$|{\cal C}| = 2*(12+498+7084+34914+96496)+146552 = 424560.$

It is shown in \cite{GrahamSloane80} that
$ A(K,4,L)  \geq \frac{1}{K}  { K \choose L }$.   This, combined with
\eqref{eq.Cbnd} and the fact  $ \sum_{L: L~\mbox{even}}  {K \choose L} = 2^{K-1},$
 implies that $\cal C$ can be selected with cardinality
at least $\frac{2^{K-1}-2}{K},$ which implies the last statement of the proposition.
\end{proofof}

\begin{rem}
The existence of speckled valuations has negative implications for the problem of
finding a computationally efficient way to present arbitrary substitute
valuations.
Suppose, for example, that an algorithm for presenting substitute
valuations  takes as input a string $x$ of real numbers, and then the algorithm determines
$v(A)$ for any bundle $A$ as a linear transformation of $x$, with coefficients depending
on $A$ and $x$.  Symbolically, we can write this as $v(A)= \sum_a  h(A,x,a) x_a$.   Suppose further
that for each bundle $A$ and index $a$, there are only finitely or countably infinitely
many possible values of the coefficient $h(A,x,a)$ as $x$ varies.
For example, the assignment valuations described in Section \ref{sec.assign_val}
can be put into this form, with the coefficients $h(A,x,a)$ taking values $\{0, 1\}$.
The same is true of the $S$-presentations and $H$-presentations given in
\cite{BingLehmannMilgrom04}.   If for every substitute valuation on $K$ items, there
is a choice of input $x$ so that the algorithm outputs the substitute valuation,
then the possible outputs of the algorithm must cover the polyhedrons in ${\cal S}_K$ consisting
of speckled valuations. 
But the set of possible output valuations is a finite or countably infinite
union of sets of dimension less than
or equal to the length of the input vector $x$.
Therefore,  in view of Proposition \ref{prop.expo_substitutes}, the length
of  $x$ must be greater than or equal to $2K-1 +  \frac{2^{K-1}-2}{K}$.
For any $\epsilon >0$, this lower bound exceeds $2^{(1-\epsilon)K}$ for sufficiently
large $K$.
\end{rem}

 \section{Discussion}
  
This paper addresses valuations for single-unit markets, for which each of the
$K$ goods is distinct.   In a multi-unit market, there may be multiple goods
of the same type, and the valuation should be invariant with respect to substituting
one good with another of the same type.   This poses additional constraints on $v$.
The Monte Carlo algorithm we presented extends immediately to this case, because if the
nominal function $\theta_o$ satisfies invariance  under swapping goods of the same
type, then the resulting $\theta$ constructed by the algorithm will be similarly invariant.
That is, to use the terminology of  \cite{MilgromStrulovici06d}, the algorithm can be
used to generate strong substitute valuations for multi-unit auctions.  Note that
by mapping from single- to mulit-unit auctions in this way, different goods of
each type can have different prices.   Another class of valuations, called weak substitute
valuations in \cite{MilgromStrulovici06d}, are defined as in Definition  \ref{def:sub}, with
the prices of all goods of the same type being the same.   It would be
interesting to find a method to generate weak substitute valuations.

It would be interesting to find an algorithm for generating substitute valuations such that
the running time is bounded by a function of $K$ alone.   As mentioned in Remark \ref{remark.large_iter},
there is no such bound for our algorithm.  Another topic for additional work is to see
how well a generation algorithm, either the one we suggested or a new one, can produce
valuations with given distributions.   Roughly speaking, if the valuations generated in
Step 1 of our algorithm have a specified distribution, and if the valuations aren't changed too
much by steps 2 and 3, then the output valuation should approximately have the given distribution.
As noted in the introduction, there is also much interest in generating realistic valuations
for various practical settings, and such valuations typically do not have the substitute property.

The examination of the richness of the class of assignment valuations and the class of
all substitute valuations in this paper is purely mathematical, rather than based on the
valuations that arise in practice.   While the existence of the speckled valuations shows that
the set of substitute valuations is much richer than the set of assignment valuations, it is
not clear whether the extra richness has practical value.   Further, there are important simple
examples of valuations which are {\em not} substitute valuations.   The most prominent of them
is the case of two complementary goods: $(v(\emptyset), v(1), v(2), v(12))=(0,0,0,1)$.

 Echenique  \cite{Echenique07} pursued a different approach to determining the
 richness of the set of substitute valuations.   The framework for his results
 is the notion of substitute introduced by  Roth  \cite{Roth84}.   Roth's framework is more
 general than the original one of \cite{KelsoCrawford82} (see \cite{HatfieldMilgrom05}).
Echenique  \cite{Echenique07}  counts the number of substitute choice functions.
The results of  \cite{Echenique07} are not directly comparable to those here, but the conclusions
are somewhat similar.

\section*{Acknowledgement}
The author is grateful to D. Lehmann, for pointing out \cite{BingLehmannMilgrom04},
and to the reviewers for helpful comments.  The work reported in this paper was
supported in part by the National Science Foundation under grant NSF ECS 06-21416.

\bibliographystyle{elsart-num-sort}

\end{document}